\let\oldvec\vec
\documentclass[runningheads]{llncs}
\let\vec\oldvec
\usepackage{subcaption}
\usepackage{wrapfig}
\usepackage{transparent}
\newcommand{\code}[1]{\texttt{#1}}
\usepackage{comment}
\usepackage{float}
\usepackage{amsmath}
\usepackage{tikz}
\usepackage[utf8]{inputenc}
\usepackage{newtxtext, newtxmath}
\usepackage{urwchancal}
\usepackage{xcolor}
\newcommand\crule[3][black]{\textcolor{#1}{\rule{#2}{#3}}}
\usepackage{color, colortbl}
\definecolor{byzantine}{rgb}{0.74, 0.2, 0.64}
\definecolor{myorange}{rgb}{1.0, 0.57, 0.0}
\definecolor{bulgarianrose}{rgb}{0.28, 0.02, 0.03}
\usepackage{graphicx}
\usepackage{adjustbox}
\usepackage{makecell}
\usepackage{wrapfig, booktabs}

\makeatletter
\newcommand*{\addFileDependency}[1]{
  \typeout{(#1)}
  \@addtofilelist{#1}
  \IfFileExists{#1}{}{\typeout{No file #1.}}
}
\makeatother
 
\newcommand*{\myexternaldocument}[1]{%
    \externaldocument{#1}%
    \addFileDependency{#1.tex}%
    \addFileDependency{#1.aux}%
}

\usepackage{xr}
\myexternaldocument{supp}

\usepackage{todonotes}

\begin{document}
\title{U-Net Fixed-Point Quantization for Medical Image Segmentation}
%
%
\author{MohammadHossein AskariHemmat\inst{1}\and
Sina Honari\inst{2} \and
Lucas Rouhier  \inst{3} \and
Christian S. Perone \inst{3} \and
Julien Cohen-Adad \inst{3} \and
Yvon Savaria \inst{1} \and
Jean-Pierre David \inst{1}}
\authorrunning{M.H. AskariHemmat et al.}
%
\institute{Ecole Polytechnique Montreal, Canada\\ \email{\{mohammadhossein.askari-hemmat, yvon.savaria, jean-pierre.david\}@polymtl.ca} 
\and Mila-University of Montreal, Canada\\ 
\email{sina.honari@umontreal.ca} 
\and NeuroPoly Lab, Institute of Biomedical Engineering, Polytechnique Montreal, Montreal 
\email{christian.perone@gmail.com, \{lucas.rouhier, julien.cohen-adad\}@polymtl.ca}}
\maketitle              
\vspace{-.2cm}
\begin{abstract}
Model quantization is leveraged to reduce the memory consumption and the computation time of deep neural networks.
This is achieved by representing weights and activations with a lower bit resolution when compared to their high precision floating point counterparts. The suitable level of quantization is directly related to the model performance. Lowering the quantization precision (e.g. 2 bits), reduces the amount of memory required to store model parameters and the amount of logic required to implement computational blocks, which contributes to reducing the power consumption of the entire system. These benefits typically come at the cost of reduced accuracy. The main challenge is to quantize a network as much as possible, while maintaining the performance accuracy. In this work, we present a quantization method for the U-Net architecture, a popular model in medical image segmentation. We then apply our quantization algorithm to three datasets: (1) the Spinal Cord Gray Matter Segmentation  (GM), (2) the ISBI challenge for segmentation of neuronal structures in Electron Microscopic (EM), and (3) the public National Institute of Health (NIH) dataset for pancreas segmentation in abdominal CT scans. The reported results demonstrate that with only 4 bits for weights and 6 bits for activations, we obtain 8 fold reduction in memory requirements while loosing only $2.21\%$,  $0.57\%$ and $2.09\%$ dice overlap score for  EM, GM and NIH datasets respectively. Our fixed point quantization provides a flexible trade-off between accuracy and memory requirement, which is not provided by previous quantization methods for U-Net. \footnote{Our code is released at \url{https://github.com/hossein1387/U-Net-Fixed-Point-Quantization-for-Medical-Image-Segmentation}}

\end{abstract}

\section{Introduction}
Image segmentation, the task of specifying the class of each pixel in an image, is one of the active research areas in the medical imaging domain. In particular, image segmentation for biomedical imaging allows identifying different tissues, biomedical structures, and organs from images to help medical doctors diagnose diseases. However, manual image segmentation is a laborious task. Deep learning methods have been used to automate the process and alleviate the burden of segmenting images manually.

The rise of Deep Learning has enabled patients to have direct access to personal health analysis \cite{10.1093/bib/bbx044}. Health monitoring apps on smartphones are now capable of monitoring medical risk factors. Medical health centers and hospitals are equipped with pre-trained models used in medical CADs to analyse MRI images \cite{thalers.menkovskiv.2019}. However, developing a high precision model often comes with various costs, such as a higher computational burden and a large model size. The latter requires many parameters to be stored in floating point precision, which demands high hardware resources to store and process images at test time. 
In medical domains, images typically have high resolution and can also be volumetric (the data has a depth in addition to width and height).
Quantizing the neural networks can reduce the feedforward computation time and most importantly the memory burden at inference. 
After quantization, a high precision (floating point) model is approximated with a lower bit resolution model. The goal is to leverage the advantages of the quantization techniques while maintaining the accuracy of the full precision floating point models.  
Quantized models can then be deployed on devices with limited memory such as cell-phones, or facilitate processing higher resolution images or bigger volumes of 3D data with the same memory budget.
Developing such methods can reduce the required memory to save model parameters potentially up to 32x in memory footprint. In addition, the amount of hardware resources (the number of logic gates) required to perform low precision computing, is much less than a full precision model \cite{DBLP:journals/corr/HubaraCSEB16}. In this paper, we propose a fixed point quantization of U-Net \cite{ronneberger2015u}, a popular segmentation architecture in the medical imaging domain. We provide comprehensive quantization results on the Spinal Cord Gray Matter Segmentation Challenge \cite{gm_orig}, the ISBI challenge for segmentation of neuronal structures in electron microscopic stacks \cite{10.1371/journal.pbio.1000502}, and the public National Institute of Health (NIH) dataset for pancreas segmentation in abdominal CT scans \cite{deeporgan.2015}. In summary, this work makes the following contributions:
\begin{itemize}
  \item We report the first fixed point quantization results on the U-Net architecture for the medical image segmentation task and show that the current quantization methods available for U-Net are not efficient for the hardware commonly available in the industry.
  \item We quantify the impact of quantizing the weights and activations on the performance of the U-Net model on three different medical imaging datasets.
  \item We report results comparable to a full precision segmentation model by using only 6 bits for activation and 4 bits for weights, effectively reducing the weights size by a factor of $8\times$ and the activation size by a factor of $5\times$.
\end{itemize}

\vspace{-5mm}
\section{Related Works}
\subsection{Image Segmentation}
Image segmentation is one of the central problems in medical imaging \cite{pham2000current}, commonly used to detect regions of interest such as tumors. Deep learning approaches have obtained the state-of-the-art results in medical image segmentation \cite{litjens2017survey,shen2017deep}. One of the favorite architectures used for image segmentation is U-Net \cite{ronneberger2015u} or its equivalent architectures proposed around the same time; ReCombinator Networks \cite{honari2016recombinator}, SegNet \cite{badrinarayanan2015segnet}, and DeconvNet \cite{noh2015learning}, all proposed to maintain pixel level information that is usually lost due to pooling layers. These models use an encoder-decoder architecture with skip connections, where the information in the encoder path is reintroduced by skip connections in the decoder path. This architecture has proved to be quite successful for many applications that require full image reconstruction while changing the modality of the data, as in the image-to-image translation \cite{isola2017image}, semantic segmentation \cite{ronneberger2015u,badrinarayanan2015segnet,noh2015learning} or landmark localization \cite{honari2016recombinator,newell2016stacked}. While all the aforementioned models propose the same architecture, for simplicity we refer to them as U-Net models. U-Net type models have been very popular in the medical imaging domain and have been also applied to the 3 dimensional (3D) segmentation task \cite{cciccek20163d}. One problem with U-Net is its high usage of memory due to full image reconstruction. All encoded features are required to be kept in memory and then used while reconstructing the final output. This approach can be quite demanding, especially for high resolution or 3D images. Quantization of weights and activations can reduce the required memory for this model, allowing to process images with a higher resolution or with a bigger 3D volume at test time.
\vspace{-5mm}
\subsection{Quantization for Medical Imaging Segmentation}
There are two approaches to quantize a neural network, namely deterministic quantization and stochastic quantization \cite{DBLP:journals/corr/HubaraCSEB16}. 
Although DNN quantization has been thoroughly studied \cite{DBLP:journals/corr/HubaraCSEB16,DBLP:journals/corr/ZhouYGXC17,NIPS20155647}, little effort has been done on developing quantization methods for medical image segmentation. In the following, we review recent works in this field.

\noindent
\textbf{Quantization in Fully Convolutional Networks}:
Quantization has been applied to Fully Convolutional Networks (FCN) in biomedical image segmentation \cite{xu2018quantization}. First, a quantization module was added to the suggestive annotation in FCN. In suggestive annotation, instead of using the original dataset, a representative training dataset was used, which in turn increased the accuracy. Next, FCN segmentations were quantized using Incremental Quantization (INQ). Authors report that suggestive annotation with INQ using 7 bits results in accuracy close to or better than those obtained with a full precision model. In FCN, features of different resolutions are upsampled back to the image resolution and merged together right before the final output predictions. This approach is sub-optimal compared to the U-Net which upsamples features only to one higher resolution, allowing the model to process them before they are passed to higher resolution layers. This gradual resolution increase in reconstruction acts as a conditional computation, where the features of higher resolution are computed using the lower resolution features. As reported in \cite{honari2016recombinator}, this process of conditional computation results in faster convergence time and increased accuracy in the U-Net type architectures compared to the FCN type architectures. Considering the aforementioned advantages of U-Net, in this paper we pursue the quantization of this model.

\noindent
\textbf{U-Net Quantization:}
  In \cite{DBLP:journals/corr/abs-1801-09449}, the authors propose the first quantization for U-Net. They introduce 1) a parameterized ternary hyperbolic tangent to be used as the activation function, 2) a ternary convolutional method that calculates matrix multiplication very efficiently in the hamming space. They report 15-fold decrease in the memory requirement as well as 10x speed-up at inference compared to the full precision model. Although this method shows significant performance boost, in Section \ref{sec:Experimental_Results} we demonstrate that this is not an efficient method for the currently available CPUs and GPUs.

\section{Proposed Quantization}
We propose fixed point quantization for U-Net. We start with a full precision (32 bit floating point) model as our baseline.
We then use the following fixed point quantization function to quantize the parameters (weights and activation) in the inference path:
\vspace{-2mm}
\begin{equation} \label{eq:quant}
\centering
 quantize(x,n) = (round(clamp(x, n) << n)) >> n
\end{equation}

where the $round$ function projects its input to the nearest integer, $<<$ and $>>$ are shift left and right operators, respectively. In our simulation, shift left and right are implemented by multiplication and division in powers of 2. The  $clamp$ function is defined as:
\vspace{-2mm}
\begin{equation} \label{eq:clamp}
\centering
clamp(x,n) = 
\begin{cases}
    2^n-1             &  $when$  \ \ \  x\geq2^n-1\\
    x                 &  $when$  \ \ \  0<x<2^n-1 \\
    0                 &  $when$  \ \ \  x \leq 0 
\end{cases}
\end{equation}

Equation (\ref{eq:quant}) quantizes an input $x\in\mathbb{R}$ to the closest value that can be represented by $n$ bits. To map any given number $x$ to its fixed point value we first split the number into its fractional and integer parts using:
\vspace{-2mm}
\begin{align}
    x_f = abs(x) - floor(abs(x)), 
    x_i = floor(abs(x)) 
\end{align}
and then use the following equation to convert $x$ to its fixed point representation using the specified number of bits for the integer ($ibits$) and fractional ($fbits$) parts:
\vspace{-2mm}
\begin{align} 
\label{eq:fixed_quant}
\centering
to\_fixed\_point(x, ibits, fbits) &= sign(x)*quantize(x_i, ibits) \nonumber\\ &+sign(x)*quantize(x_f, fbits)
\end{align}

Equation (\ref{eq:fixed_quant})  is a fixed point quantization function that maps a floating point number $x$ to the closest fixed point value with $ibits$ integer and $fbits$ fractional bits. Throughout this paper, we use $Q^{p}i.f$ notation to denote that we are using a fixed point quantization of parameter $p$ by using $i$ bits to represent the integer part and $f$ bits to represent the fractional part. Based on our experiments, we did not benefit from an incremental quantization (INQ) as explained in \cite{DBLP:journals/corr/ZhouYGXC17}. Although this method could work for higher precision models, for instance when using fixed point $Q^w8.8$ (Quantizing weights with 8-bit integer and 8-bit fractional parts), for extreme quantization as in $Q^w0.4$, learning from scratch gave us the best accuracy with the shortest learning time.
As shown in Figure \ref{fig:quant_network}, in the full precision case, the weights of all U-Net layers are in $[-1, 1]$ range, hence the integer part for the weight quantization is not required. 

\subsection{Training}
For numerical stability and to verify the gradients can propagate in training, we demonstrate that our quantization is differentiable . Starting from Equation (\ref{eq:clamp}), the derivative is:
\vspace{-3mm}
\begin{equation} \label{eq:clamp_der}
\centering
\forall x\in\mathbb{R}, \forall n \in \mathbb{Z^+},  \ \ \  \frac{\partial}{\partial x}clamp(x,n) = 
\begin{cases}
    0                 &  $when$  \ \ \  x\geq2^n-1\\
    1                 &  $when$  \ \ \  0<x<2^n-1 \\
    0                 &  $when$  \ \ \  x \leq 0 
\end{cases}
\end{equation}
\noindent

which is differentiable except on the thresholds. To make it completely differentiable, 
a straight-through estimator (STE), introduced in \cite{STECoursera}, is used. The STE passes gradients over the thresholds and also over the $round$ function in Equation (\ref{eq:quant}).  

\subsection{Observations on U-Net Quantization}
\subsubsection{Dropout}
Dropout \cite{srivastava2014dropout} is a regularization technique to prevent over-fitting of DNNs. Although it is used in the original implementation of U-Net, we found that when this technique is applied along with quantization, the accuracy drops a lot. Hence, in our implementation, we removed dropout from all layers.
This is due to the fact that quantization acts as a strong regularizer, as reported in  \cite{DBLP:journals/corr/HubaraCSEB16}, hence further regularization with dropout is not required. 
As shown in Figure \ref{fig:dropout}, for each quantized precision, dropout reduces the accuracy, with the gap being even higher for lower precision quantizations.

\vspace{-.5cm}
\subsubsection{Full Precision Layers}
It is well accepted to keep the first and the last layers in full precision, when applying quantization \cite{DBLP:journals/corr/HubaraCSEB16,Tang2017HowTT}. 
However, we found that in the segmentation task, keeping the last layer in full precision has much more impact than keeping the first layer in full precision. 
\vspace{-.5cm}

\subsubsection{Batch Normalization}

Batch normalization is a technique that improves the training speed and accuracy of DNN. 
We used the Pytorch implementation of batchnorm. In training, we use the quantization block after the batchnorm block in each layer 
such that the batchnorm is first applied using the floating point calculations and then the quantized value is sent to the next layer (hence not quantizing the batchnorm block during training). However, at inference, Pytorch folds the batchnorm parameters into the weights, effectively including batchnorm parameters in the quantized model as part of the quantized weights.
\section{ Results and Discussion}
\label{sec:Experimental_Results}
\begin{figure}[!t]
  \centering
    \includegraphics[width=1.0\textwidth]{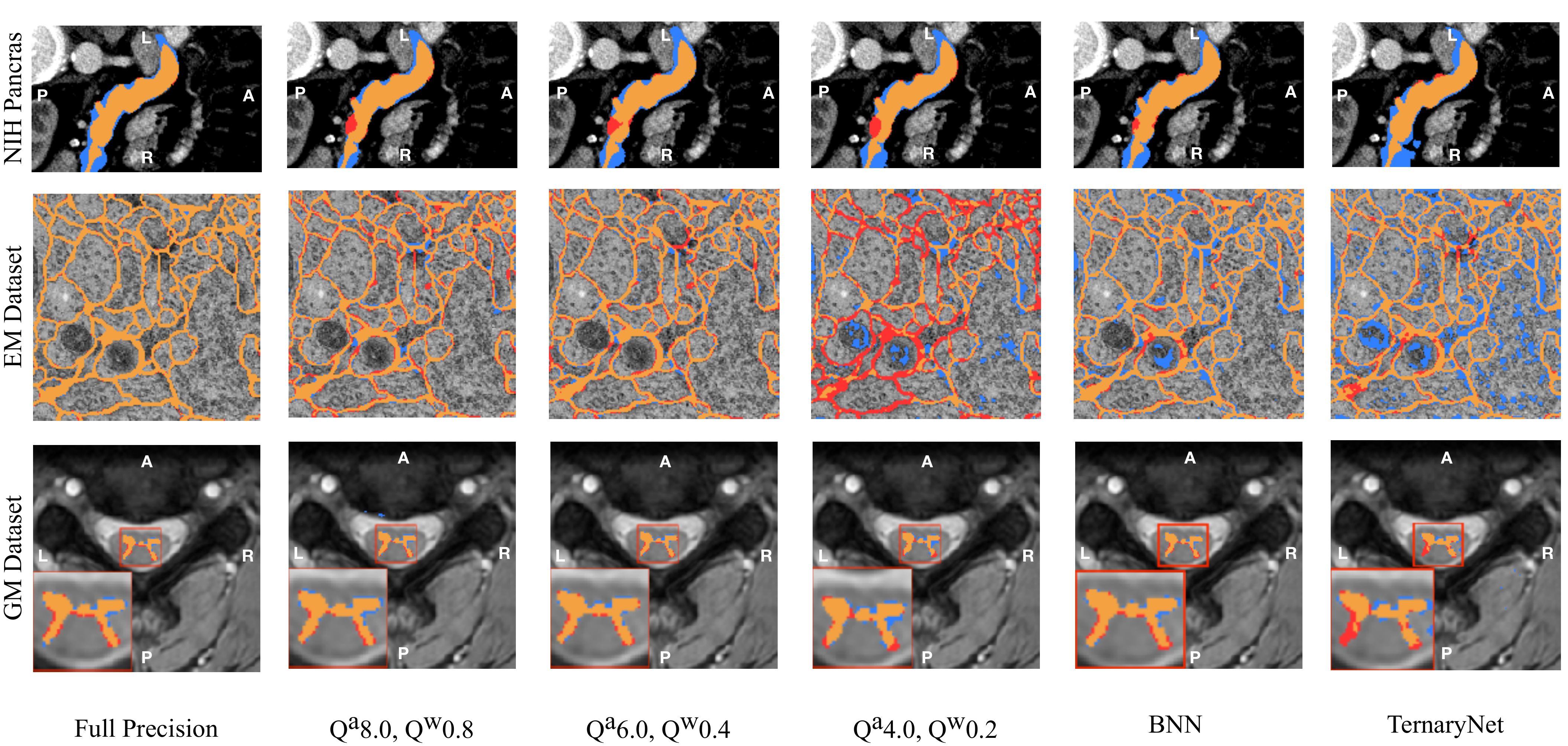}
    \small
\caption{\small Sample prediction versus ground truth segmentation results for NIH Pancreas (top), EM (middle) and GM (bottom) datasets.  From left to right, the result of different quantization methods and precisions are reported. 
Segments in \crule[blue]{0.2cm}{0.2cm} show false positive, segments in  \crule[red]{0.2cm}{0.2cm} show false negative and segments in \crule[myorange]{0.2cm}{0.2cm} show true positive. 
}
\label{fig:main}
\vspace{-2mm}
\end{figure}
We implemented the U-Net model and our fixed-point quantizer in Pytorch. 
We trained our model over 200 epochs with a batch size of 4.
We applied our fixed point quantization along with TernaryNet \cite{DBLP:journals/corr/abs-1801-09449} and Binary \cite{NIPS20155647} quantization on three different datasets: GM \cite{gm_orig}, EM \cite{10.1371/journal.pbio.1000502}, and NIH \cite{deeporgan.2015}. For GM and EM datasets, we used an initial learning rate of $1e-3$, and for NIH we used initial learning rate of $0.0025$. For all datasets we used Glorot for weight initialization and cosine annealing scheduler to reduce learning rate in training. Please check our repository for the model and training details.
\footnote{\url{https://github.com/hossein1387/U-Net-Fixed-Point-Quantization-for-Medical-Image-Segmentation}}

The NIH pancreas \cite{deeporgan.2015} dataset is composed of 82 3D abdominal CT scan and their corresponding pancreas segmentation images. Unfortunately, we did not have access to the pre-processed dataset described in \cite{DBLP:journals/corr/abs-1801-09449}, nevertheless, we extracted 512x512 2-D slices from the original dataset and applied a region of interest cropping to get 7059 images of size 176x112. The final dataset contains 7059 176x112 2-D images which are separated into training and testing dataset (respectively 80\% and 20\%). For GM and EM datasets, we used the provided dataset as described in \cite{gm_orig} and \cite{10.1371/journal.pbio.1000502} respectively. For both EM and GM datasets, we did not used any region of interest cropping and we used images of size 200x200. 

The task of image segmentation for GM and NIH pancreas datasets is imbalanced. As suggested in \cite{gm_orig}, instead of weighted cross-entropy, we used a surrogate loss for the dice similarity coefficient. This loss is referred to as  the dice loss and is formulated as 
$ \mathcal{L}_{dice} = \frac{2\sum^N_{n=1}p_nr_n +\epsilon}{\sum^N_{n=1}p_n +\sum^N_{n=1}r_n + \epsilon}$,
where $p_n\in[0, 1]$ and $r_n\in\{0, 1\}$ are prediction and ground truth pixels respectively (with $0$ indicating \textit{not-belonging} and $1$ indicating \textit{belonging} to the class of interest) and $\epsilon$ is the noise added for numerical stability. For the EM dataset, using a weighted sum of cross entropy and dice loss produced the best results. 

Figure \ref{fig:main} along with Table \ref{table:scores} show the impact of different quantization methods on the aforementioned datasets. Considering the NIH dataset, Figure \ref{fig:main}(top) and Table \ref{table:scores} show that despite using only 1 and 2 bits to represent network parameters, Binary and TernaryNet quantizations produce results that are close to the full precision model. However, for other datasets, our fixed point $Q^a$6.0, $Q^w$0.4 quantization surpasses Binary and TernaryNet quantization. The other important factor here is how efficient these quantization techniques can be implemented using the current CPUs and GPUs hardware. At the time of writing this paper, there is no commercially available CPU or GPU that can efficiently store and load sub-8-bit parameters of a neural network, which leaves us to use custom functions to do bit manipulation to make sub-8-bit quantization more efficient. Moreover, in the case of TernaryNet, to apply floating point scaling factors after ternary convolutions, floating point operations are required. Our fixed point quantization uses only integer operations, which requires less hardware footprint and use less power compared to floating point operations. Finally, TernaryNet uses Tanh instead of ReLU for the activations. 
Using hyperbolic tangent as an activation function increases training time \cite{NIPS2012_4824} and execution time at inference. To verify it, we evaluated the performance of ReLU and Tanh in a simple neural network with 3 fully connected layers. We used the Intel's OpenVino \cite{deannedeuermeyerandreyz.amyr.fritzb.2019} inference engine together with high performance \code{gemm\_blas} and \code{avx2} instructions. Table \ref{table:relu_vs_tanh} reports the results obtained when ReLU is used instead of Tanh at training and it shows that inference time can decrease by up to 8 times. These results can be extended to U-Net, since activation inference time is a direct function of the input size.
To compensate for the computation time, TernaryNet implements an efficient ternary convolution that can decrease processing time by up to 8 times. At inference, an efficient Tanh function that uses only two comparators can be implemented to perform Tanh for ternary values. 
Considering accuracy, when Tanh is used as an activation function, the full precision accuracy is lower compared to ReLU \cite{DBLP:journals/corr/abs-1801-09449}. We observe similar behavior in the results reported in Table \ref{table:scores}. Our fixed point quantizer provides a flexible trade-off between accuracy and memory, which makes it a practical solution for the current CPUs and GPUs, as it does not require floating-point operations, and leverages the more efficient ReLU function. As opposed to BNN and TernaryNet quantizations, Table \ref{table:scores} shows that our approach for quantization of U-Net provides consistent results over 3 different datasets.

\begin{table}[!t]
\caption{\small Dice scores of the quantized U-Net models on EM (left) GM (middle) and NIH (right) datasets. The last two rows show results for Binary and TernaryNet quantizations. Other rows report results obtained for different weights and activations quantization precisions. For the GM and EM datasets, we also report results when Tanh is used instead of ReLU as the activation function.}
\centering
\resizebox{.85\linewidth}{!}{
\begin{tabular}{|c|c|l|l|l|l|l|l|}
\hline
\multicolumn{2}{|c|}{Quantization} &  & \multicolumn{2}{c|}{EM Dataset} & \multicolumn{2}{c|}{GM Dataset} & \multicolumn{1}{c|}{NIH Panceas} \\ \hline
Activation & Weight & \multicolumn{1}{c|}{\begin{tabular}[c]{@{}c@{}}Parameter\\ Size\end{tabular}} & \multicolumn{1}{c|}{\begin{tabular}[c]{@{}c@{}}Dice Score\\ ReLU\end{tabular}} & \multicolumn{1}{c|}{\begin{tabular}[c]{@{}c@{}}Dice Score\\ Tanh\end{tabular}} & \multicolumn{1}{c|}{\begin{tabular}[c]{@{}c@{}}Dice Score\\ ReLU\end{tabular}} & \multicolumn{1}{c|}{\begin{tabular}[c]{@{}c@{}}Dice Score\\ Tanh\end{tabular}} & \multicolumn{1}{c|}{Dice Score} \\ \hline
\multicolumn{2}{|c|}{Full Precision} & 18.48 MBytes & 94.05 & 93.02 & 56.32 & 56.26 & 75.69 \\ \hline
Q8.8 & Q8.8 & 9.23 MBytes & 92.02 & 91.08 & 56.11 & 56.01 & 74.61 \\ \hline
Q8.0 & Q0.8 & 4.61 MBytes & 92.21 & 88.42 & 56.10 & 53.78 & 73.05 \\ \hline
Q6.0 & Q0.4 & 2.31 MBytes & 91.03 & 90.93 & 55.85 & 52.34 & 73.48 \\ \hline
Q4.0 & Q0.2 & 1.15 MBytes & 79.80 & 54.23 & 51.80 & 48.23 & 71.77 \\ \hline
\multicolumn{2}{|c|}{BNN \cite{NIPS20155647}} & 0.56 MBytes & 78.53 & - & 31.44 & - & 72.56 \\ \hline
\multicolumn{2}{|c|}{TernaryNet \cite{DBLP:journals/corr/abs-1801-09449}} & 1.15 MBytes & - & 82.66 & - & 43.02 & 73.9 \\ \hline
\end{tabular}
} 
\label{table:scores}
\end{table}

\vspace{-25pt}
\begin{table}
\caption{\small 
Comparing ReLU and Tanh run time using Intel's OpenVino \cite{deannedeuermeyerandreyz.amyr.fritzb.2019}. Each row illustrates the execution time for a layer of a neural network in micro seconds. It demonstrates that using Tanh as activation can increase execution time by up to 8 times compared to ReLU. }
\begin{adjustbox}{width=12cm,center}
\begin{tabular}{lccccc}
\hline 
Layer Type & Instruction Type & \makecell{Execution time \\ in $\mu$s Tanh }&  \makecell{Execution time \\ in $\mu$s ReLU} & \makecell{Performance Gain of \\using ReLU over Tanh }& Tensor Dimension \\ \hline
\code{Activation} & \code{jit\_avx2\_FP32}      & \code{30 } & \code{5  } & \code{6  } & \code{[100, 100]  } \\ 
\code{FullyConnected} & \code{gemm\_blas\_FP32} & \code{20 } & \code{19 } & \code{-  } & \code{-  } \\ 
\code{FullyConnected} & \code{gemm\_blas\_FP32} & \code{860} & \code{527} & \code{-  } & \code{-  } \\ 
\code{Activation} & \code{jit\_avx2\_FP32}      & \code{77 } & \code{9  } & \code{8.6  } & \code{[100, 300]  } \\ \hline
\end{tabular}
\end{adjustbox}
\label{table:relu_vs_tanh}
\end{table}

\vspace{-30pt}

\section{Conclusion}
In this work, we proposed a fixed point quantization method for the U-Net architecture and evaluated it on the medical image segmentation task. We reported quantization results on three different segmentation datasets and showed that our fixed point quantization produces more accurate and also more consistent results over all these datasets compared to other quantization techniques. We also demonstrated that Tanh, as the activation function, reduces the base-line accuracy and also adds a computational complexity in both training and inference. Our proposed fixed-point quantization technique provides a trade-off between accuracy and the required memory. It does not require floating-point computation and it is more suitable for the currently available CPUs and GPUs hardware. 


{\small
\bibliographystyle{splncs-nopagenum}
\bibliography{references}
}

\clearpage

 



\renewcommand{\thesection}{S.\arabic{section}}
 \setcounter{section}{0}
 \renewcommand{\thesubsection}{\thesection.\arabic{subsection}}

\newcommand{\beginsupplementary}{%
   \setcounter{table}{0}
   \renewcommand{\thetable}{S\arabic{table}}%
   \setcounter{figure}{0}
   \renewcommand{\thefigure}{S\arabic{figure}}%
 }

\beginsupplementary

\onecolumn

{%
\begin{center}
\textbf{\Large Supplementary Information for \\
 U-Net Fixed-Point Quantization for Medical Image Segmentation}
\vspace{1. em} \nonumber
\end{center}
}

%
%






%
%
%
\section{Weight Visualization of Full-Precision U-Net}
\label{}
\begin{figure}[!th]
\centering
  \begin{minipage}{.6\textwidth}
    \centering
    \includegraphics[width=\linewidth]{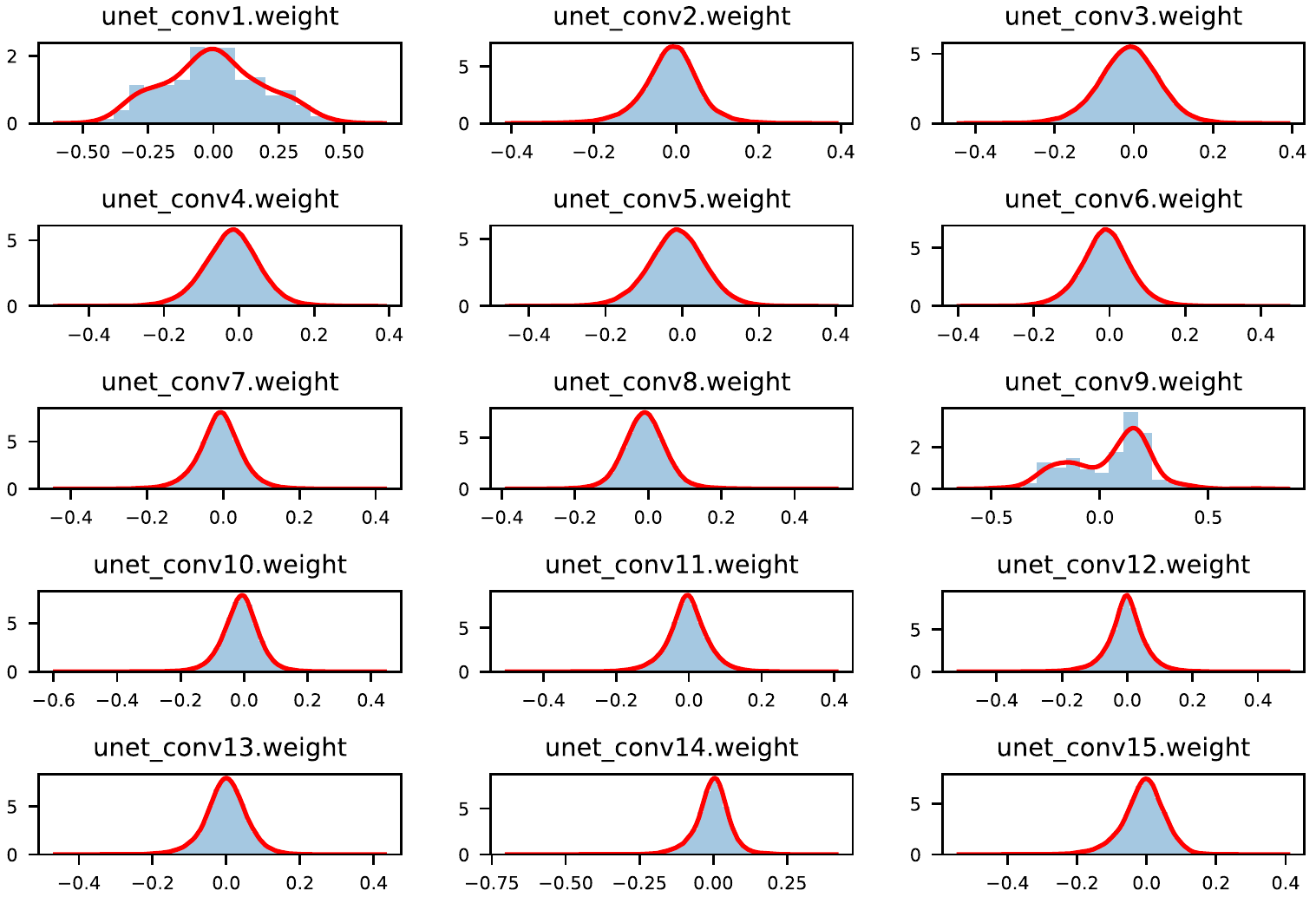}
  \end{minipage}\vspace{0.2cm}
  \begin{minipage}{.6\textwidth}
  \end{minipage}
\caption{
The weight distribution of U-Net when no quantization is used (full precision). 
The red curve shows the kernel density estimate (KDE) of the weights and the blue bins show the actual weight values of the distribution. Each plot shows the KDE of one convolutional layer in U-Net. As observed in this figure, in the full precision model almost all weights are distributed in $[-1, +1]$. Hence, no integer part is required in the quantized weights and using higher number of bits for the fractional part of the weights will result in a higher precision.}
\label{fig:quant_network}
\end{figure}

\section{Model Architecture}
\label{supp:Model_Architecture}
\small{
\begin{verbatim}
----------------------------------------------------------------
        Layer (type)          Output Shape           Param #
================================================================
            Conv2d-1        [ 64 , 200, 200]             640
       BatchNorm2d-2        [ 64 , 200, 200]             128
        QuantLayer-3        [ 64 , 200, 200]               0
            Conv2d-4        [ 64 , 200, 200]          36,928
       BatchNorm2d-5        [ 64 , 200, 200]             128
        QuantLayer-6        [ 64 , 200, 200]               0
          DownConv-7        [ 64 , 200, 200]               0
         MaxPool2d-8        [ 64 , 100, 100]               0
       Conv2dQuant-9        [ 128, 100, 100]          73,856
      BatchNorm2d-10        [ 128, 100, 100]             256
       QuantLayer-11        [ 128, 100, 100]               0
      Conv2dQuant-12        [ 128, 100, 100]         147,584
      BatchNorm2d-13        [ 128, 100, 100]             256
       QuantLayer-14        [ 128, 100, 100]               0
         DownConv-15        [ 128, 100, 100]               0
        MaxPool2d-16        [ 128, 50 , 50 ]               0
      Conv2dQuant-17        [ 256, 50 , 50 ]         295,168
      BatchNorm2d-18        [ 256, 50 , 50 ]             512
       QuantLayer-19        [ 256, 50 , 50 ]               0
      Conv2dQuant-20        [ 256, 50 , 50 ]         590,080
      BatchNorm2d-21        [ 256, 50 , 50 ]             512
       QuantLayer-22        [ 256, 50 , 50 ]               0
         DownConv-23        [ 256, 50 , 50 ]               0
        MaxPool2d-24        [ 256, 25 , 25 ]               0
      Conv2dQuant-25        [ 256, 25 , 25 ]         590,080
      BatchNorm2d-26        [ 256, 25 , 25 ]             512
       QuantLayer-27        [ 256, 25 , 25 ]               0
      Conv2dQuant-28        [ 256, 25 , 25 ]         590,080
      BatchNorm2d-29        [ 256, 25 , 25 ]             512
       QuantLayer-30        [ 256, 25 , 25 ]               0
         DownConv-31        [ 256, 25 , 25 ]               0
         Upsample-32        [ 256, 50 , 50 ]               0
      Conv2dQuant-33        [ 256, 50 , 50 ]       1,179,904
      BatchNorm2d-34        [ 256, 50 , 50 ]             512
       QuantLayer-35        [ 256, 50 , 50 ]               0
      Conv2dQuant-36        [ 256, 50 , 50 ]         590,080
      BatchNorm2d-37        [ 256, 50 , 50 ]             512
       QuantLayer-38        [ 256, 50 , 50 ]               0
         DownConv-39        [ 256, 50 , 50 ]               0
           UpConv-40        [ 256, 50 , 50 ]               0
         Upsample-41        [ 256, 100, 100]               0
      Conv2dQuant-42        [ 128, 100, 100]         442,496
      BatchNorm2d-43        [ 128, 100, 100]             256
       QuantLayer-44        [ 128, 100, 100]               0
      Conv2dQuant-45        [ 128, 100, 100]         147,584
      BatchNorm2d-46        [ 128, 100, 100]             256
       QuantLayer-47        [ 128, 100, 100]               0
         DownConv-48        [ 128, 100, 100]               0
           UpConv-49        [ 128, 100, 100]               0
         Upsample-50        [ 128, 200, 200]               0
      Conv2dQuant-51        [ 64 , 200, 200]         110,656
      BatchNorm2d-52        [ 64 , 200, 200]             128
       QuantLayer-53        [ 64 , 200, 200]               0
      Conv2dQuant-54        [ 64 , 200, 200]          36,928
      BatchNorm2d-55        [ 64 , 200, 200]             128
       QuantLayer-56        [ 64 , 200, 200]               0
         DownConv-57        [ 64 , 200, 200]               0
           UpConv-58        [ 64 , 200, 200]               0
           Conv2d-59        [ 1  , 200, 200]             577
================================================================
Total params: 4,837,249
Trainable params: 4,837,249
Non-trainable params: 0
----------------------------------------------------------------
Input size (MB): 0.15
Forward/backward pass size (MB): 593.57
Params size (MB): 18.45
Estimated Total Size (MB): 612.17
----------------------------------------------------------------\end{verbatim}
}

\section{Dropout In Quantization}
\label{supp:Dropout_Quantization}
\begin{figure}[!h]
  \centering
  \includegraphics[width=9.cm]{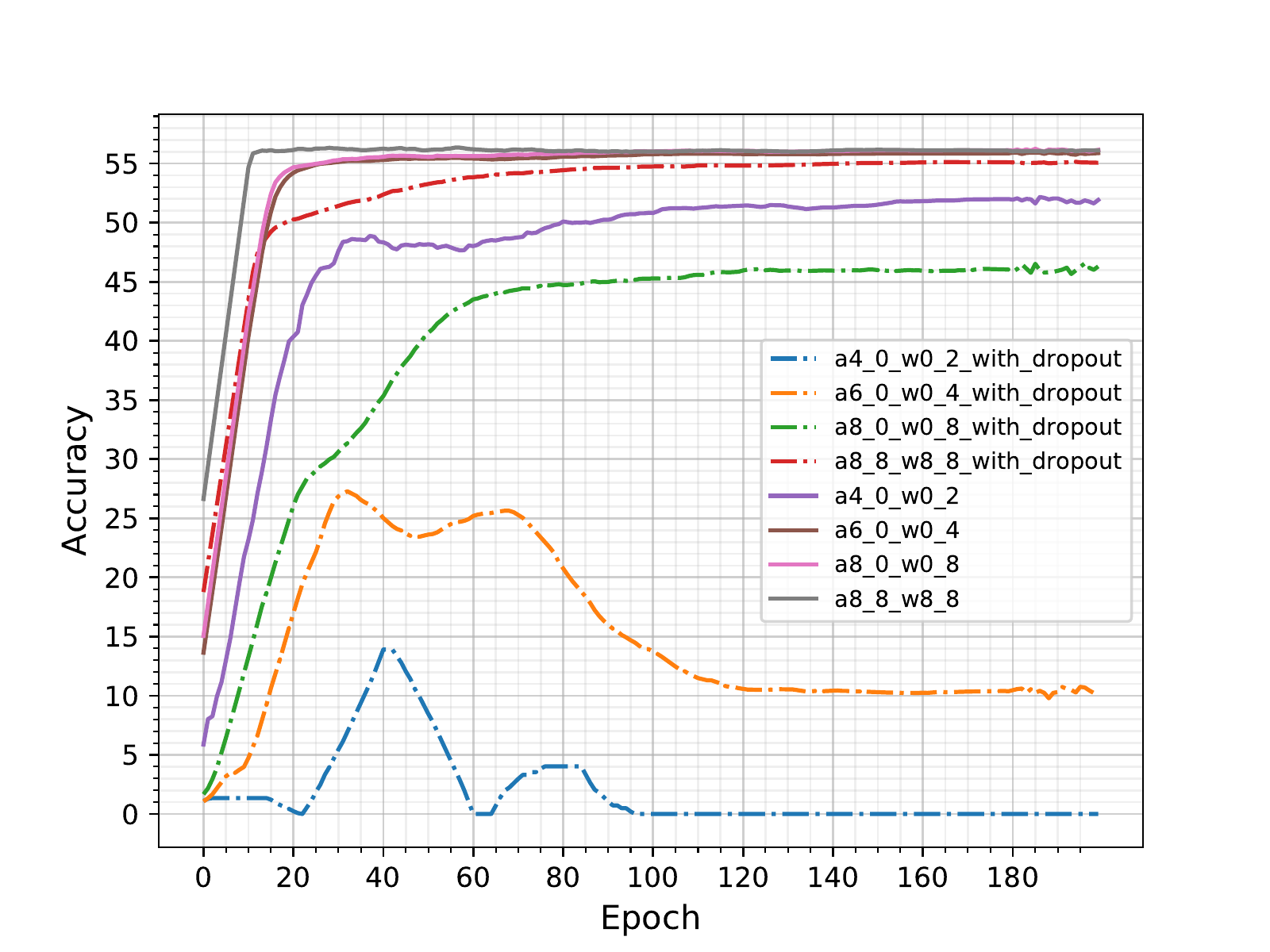}
\caption{Dice score of different U-Net quantization precisions over 200 epochs on the Spinal Cord Gray Matter Segmentation data set. We used our fixed point quantizer to show that using drop out while applying quantization on the model can drastically reduce the dice score. This phenomenon has more severe impact when the model precision is reduced. Curves shown with $--$ use dropout and quantization.}
\label{fig:DICE_WITH_WITHOUT_DROPOUT}
\label{fig:dropout}
\vspace{-5mm}
\end{figure}

\newpage
\section{More Experimental Results}
\label{supp:More_results}
\subsection{More Experimental Results For EM dataset}
\begin{figure}[!h]
  \centering
    \includegraphics[width=\textwidth]{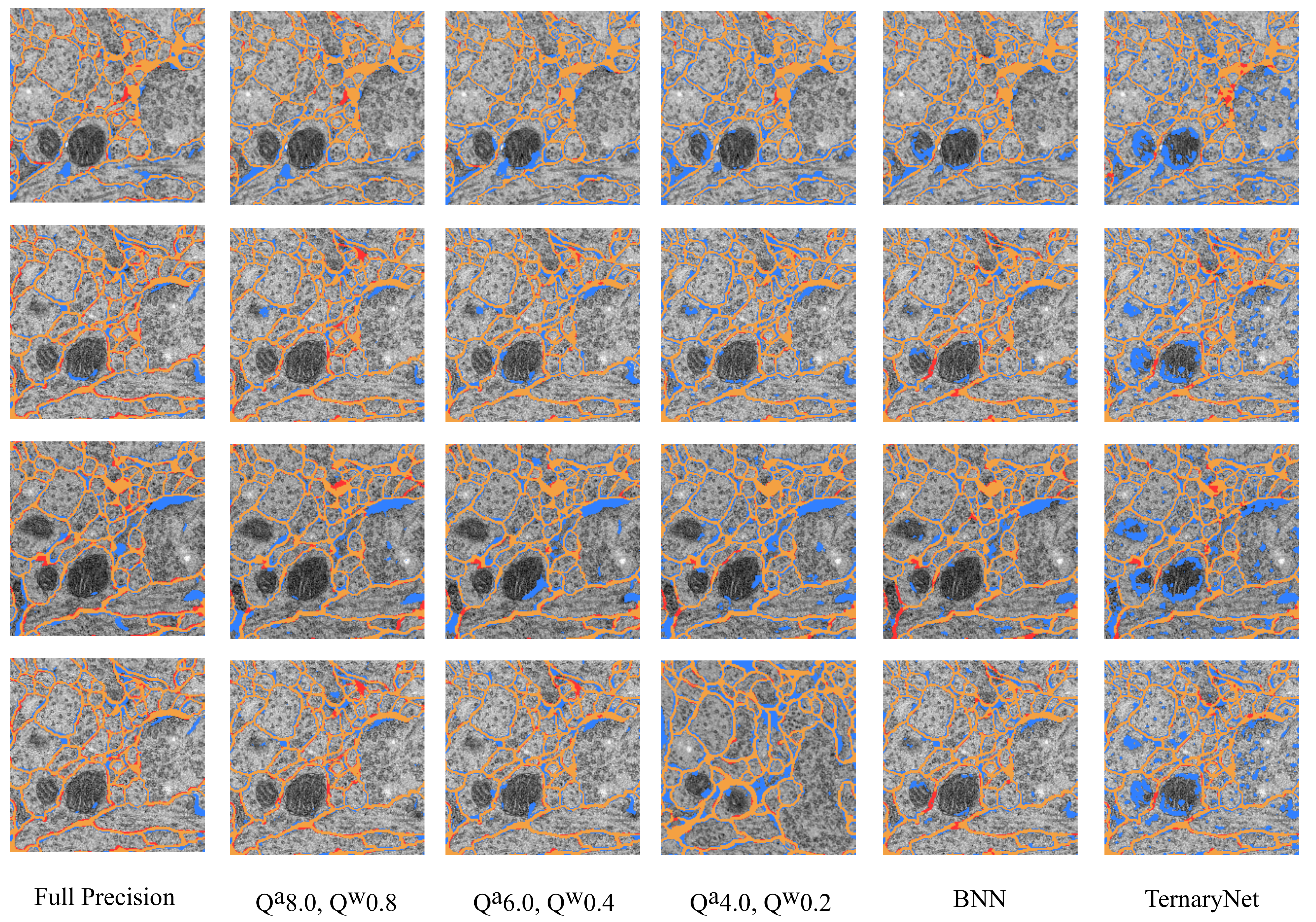}
    \small
\caption{\small Sample prediction versus ground truth segmentation results for EM data set.  From left to right, results for different quantization precision are reported. 
Segments in \crule[blue]{0.2cm}{0.2cm} show false positive, segments in  \crule[red]{0.2cm}{0.2cm} show false negative and segments in \crule[myorange]{0.2cm}{0.2cm} show true positive. The quantized model of $Q^a6.0$, $Q^w0.4$ obtains close results compared to the full precision model.
}
\label{fig:More_EM_Results}
\vspace{-2mm}
\end{figure}

\newpage
\subsection{More Experimental Results For NIH Pancreas dataset}
\begin{figure}[!h]
  \centering
    \includegraphics[width=\textwidth]{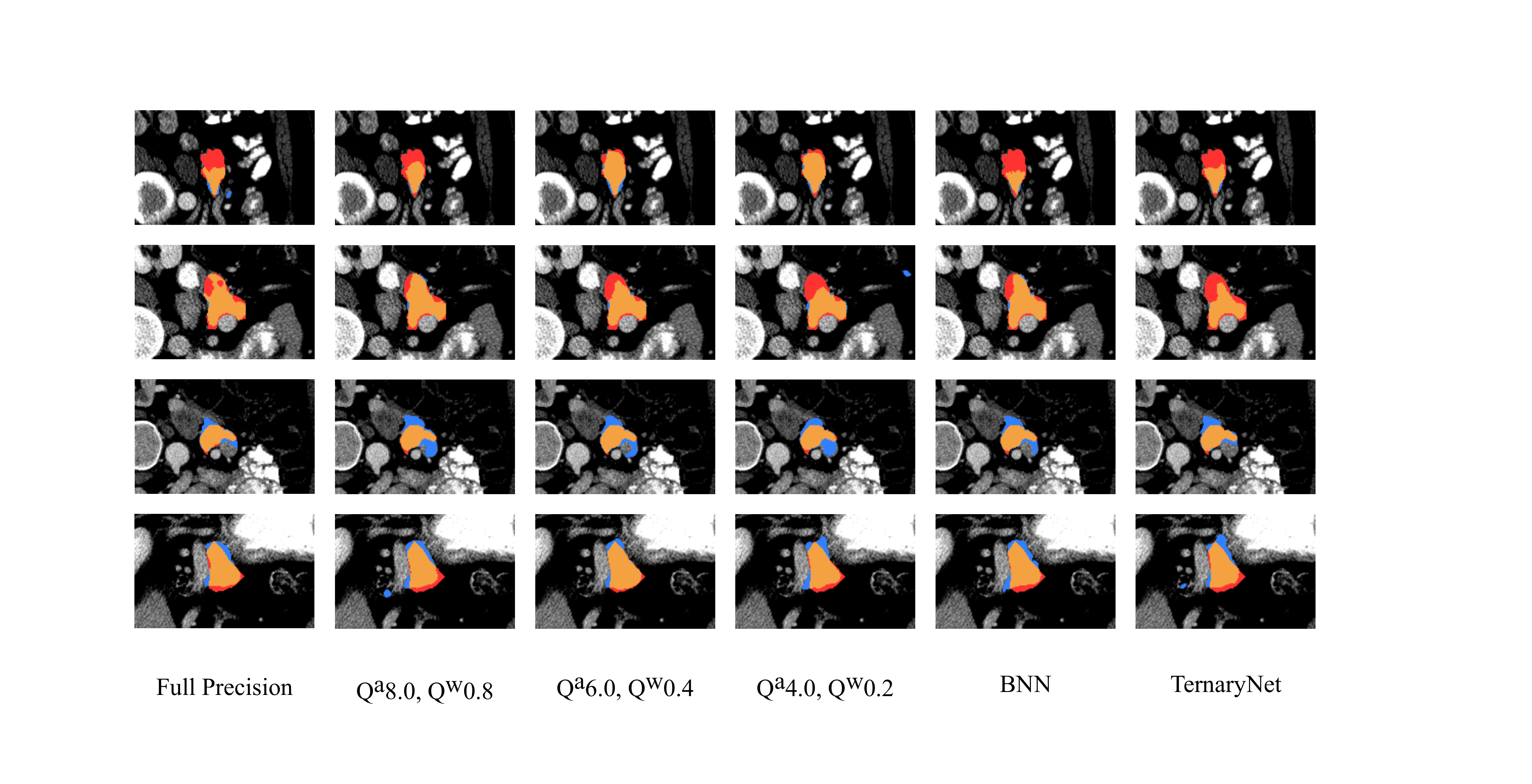}
    \small
\caption{\small Sample prediction versus ground truth segmentation results for NIH pancreas dataset.  From left to right, results for different quantization methods. 
Segments in \crule[blue]{0.2cm}{0.2cm} show false positive, segments in  \crule[red]{0.2cm}{0.2cm} show false negative and segments in \crule[myorange]{0.2cm}{0.2cm} show true positive. Note that in contrary to GM dataset, NIH pancreas segmentation is much harder since the pancreas has a very high anatomical variability. }
\label{fig:More_EM_Results}
\vspace{-2mm}
\end{figure}

\newpage
\subsection{More Experimental Results For GM dataset}
\begin{figure}[!h]
  \centering
    \includegraphics[width=\textwidth]{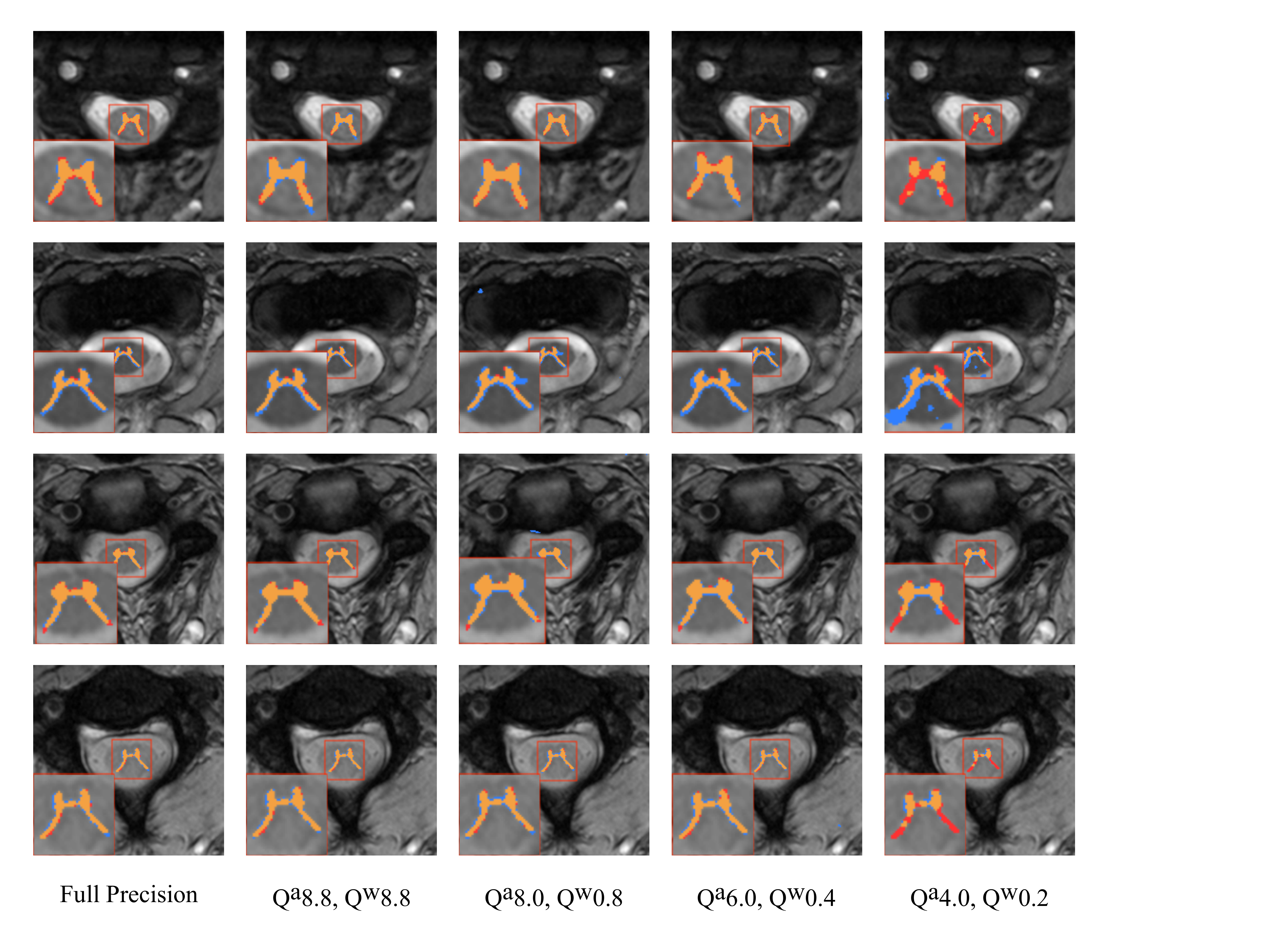}
    \small
\caption{\small Sample prediction versus ground truth segmentation results for GM data set.  From left to right, results for different quantization precision are reported. 
Segments in \crule[blue]{0.2cm}{0.2cm} show false positive, segments in  \crule[red]{0.2cm}{0.2cm} show false negative and segments in \crule[myorange]{0.2cm}{0.2cm} show true positive. The quantized model of $Q^a6.0$, $Q^w0.4$ obtains close results compared to the full precision model.
}
\label{fig:More_GM_Results}
\vspace{-2mm}
\end{figure}

\end{document}